\documentclass[a4paper]{PoS}

\newcommand{\met}{\ensuremath{{\not\mathrel{E}}_T}~}
\def\pt{$p_{T}$~}
\def\fb{~fb$^{-1}$~}

\def\W{$W$~}
\def\Z{$Z$~}
\def\pz{$p_{Z}$~}

\title{Measurements of $W$ charge asymmetry}

\ShortTitle{Measurements of $W$ charge asymmetry}

\author{\speaker{J. L. Holzbauer}\thanks{On behalf of the D0 collaboration.}\\
        University of Mississippi\\
        E-mail: \email{jholzbau@fnal.gov}}

\abstract{We discuss \W boson and lepton charge asymmetry measurements from \W decays in the electron channel, which were made using 9.7\fb of RunII data collected by the D0 detector at the Fermilab Tevatron Collider. The electron charge asymmetry is presented as a function of pseudo-rapidity out to |$\eta$| $\le$ 3.2, in five symmetric and asymmetric kinematic bins of electron transverse momentum and the missing transverse energy of the event.  We also give the \W charge asymmetry as a function of \W boson rapidity. The asymmetries are compared with next-to-leading order perturbative quantum chromodynamics calculations. These charge asymmetry measurements will allow more accurate determinations of the proton parton distribution functions and are the most precise to date.}

\FullConference{The European Physical Society Conference on High Energy Physics\\
		22--29 July 2015\\
		Vienna, Austria}

\begin{document}
The structure of the proton, described via parton distribution functions (PDFs), is of particular interest in the modern particle physics era of large proton colliders.  One important measurement used to formulate PDFs is the \W charge asymmetry, the difference in the number of positively and negatively charged \W bosons over the sum as a function of \W boson production angle.  However, complications arise because the \W boson itself cannot be fully reconstructed directly at hadron colliders due to the inaccessibility of the neutrino longitudinal momentum (\pz) information.  

\section{Methods}
There are two methods used here to determine the \W asymmetry.  One, the traditional method, is to measure the lepton charge asymmetry in \W events.  In this method, the \textit{V-A} structure of the \W boson decay modifies the asymmetry and increases the uncertainty relative to a hypothetical measurement using the full \W boson information, particularly in forward lepton pseudorapidity ($\eta$) regions.  The other method~\cite{wasymmethod} determines the \W charge asymmetry by reconstructing the \W boson assuming the \W boson mass and using other event information to estimate the likely neutrino \pz value.  The uncertainty for this measurement is lower than the first method, as more information about the event is used and the \W is accessed more directly.  These two methods will be referred to as the lepton method and the reconstructed \W method in this work. 

\section{Detector}
Data used in this analysis originated from 1.96 TeV $p\overline{p}$ collisions produced by the Fermilab Tevatron Collider.  The data were collected by the D0 detector~\cite{d0:overview}, which is a multi-purpose detector with inner tracker, calorimeter and muon systems, and is described in more detail elsewhere.  Operations ended in September of 2011 and this analysis uses the full 9.7\fb data set.  

The analysis benefits from certain special aspects of the detector and collider.  The studies involve measuring the charge of the electron, so they benefit from regular reversals of magnet polarity and the symmetric nature of the D0 detector.  More generally, because the colliding particles are $p$ and $\overline{p}$, the initial states are CP symmetric, allowing the positive and negative rapidity regions in the analysis to be combined to increase statistics.  Additionally, the \W is largely formed from valence quarks, rather than the sea quarks and gluons (which form most \W's at the CERN Large Hadron Collider), allowing this study to provide different information than one done using $pp$ collisions.  

The analysis discussed in this proceedings, done in the electron channel, is documented in greater detail elsewhere for both the reconstructed \W~\cite{D0:wasym} and lepton~\cite{D0:lepasym} methods.  Additionally, studies of \W charge asymmetry in both muon and electron channels have been performed previously by D0~\cite{D0:muonasym}, CDF~\cite{CDF:lepasym, CDF:wasym}, ATLAS~\cite{ATLAS:wasym}, CMS~\cite{CMS:wasym1, CMS:wasym2} and LHCb~\cite{lhcb:wasym}.  The analysis reported here using the lepton method improves upon and replaces the previous lepton method D0 electron channel \W asymmetry result~\cite{D0:oldasym}, which was done with 1\fb of data.

\section{Analysis Selection and Backgrounds}
In the case of both methods, the analysis selections are the same.  Exactly one electron is required, and this electron must be triggered, isolated, have most of its energy contained within the electro-magnetic calorimeter, and the calorimeter cluster must have a track matched to it.  The electron is required to have an |$\eta$| < 1.1 or 1.5 < |$\eta$| < 3.2.  The electron \pt and the missing transverse energy (\met) are both required to be above 25 GeV, and the electron \pt must also be below 100 GeV to ensure reasonable track (and charge) resolution.  Additionally, there are event quality requirements, including restrictions of the z vertex range, \W boson transverse mass, recoil and total calorimeter activity.

The primary backgrounds for this analysis are $W \rightarrow \tau \nu$, $Z \rightarrow ee$, $Z  \rightarrow \tau \tau$, and QCD.  The largest is QCD at 4 percent, although the impact on the analysis is minimal because QCD does not have an inherent charge asymmetry.  

\section{Efficiencies and Corrections}
This analysis has several efficiencies and corrections, including charge mis-ID, electron energy scale, trigger, hadronic response, electron ID efficiency, etc.  For more detail, please consult the full PRD on this topic.  

\begin{figure}
\begin{minipage}{0.50\linewidth}
\centerline{\includegraphics[width=0.9\linewidth]{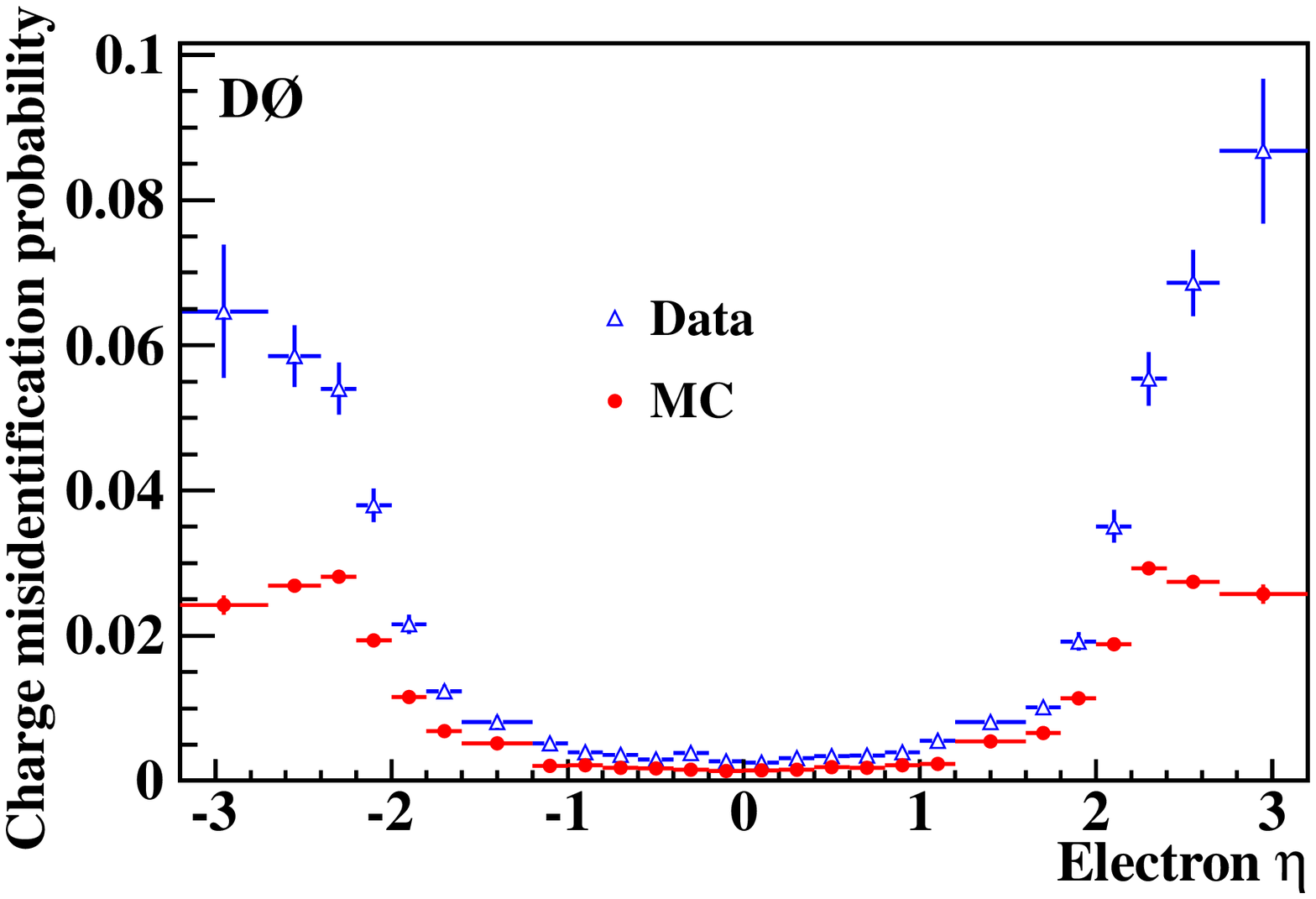}}
\end{minipage}
\hfill
\begin{minipage}{0.50\linewidth}
\centerline{\includegraphics[width=0.9\linewidth]{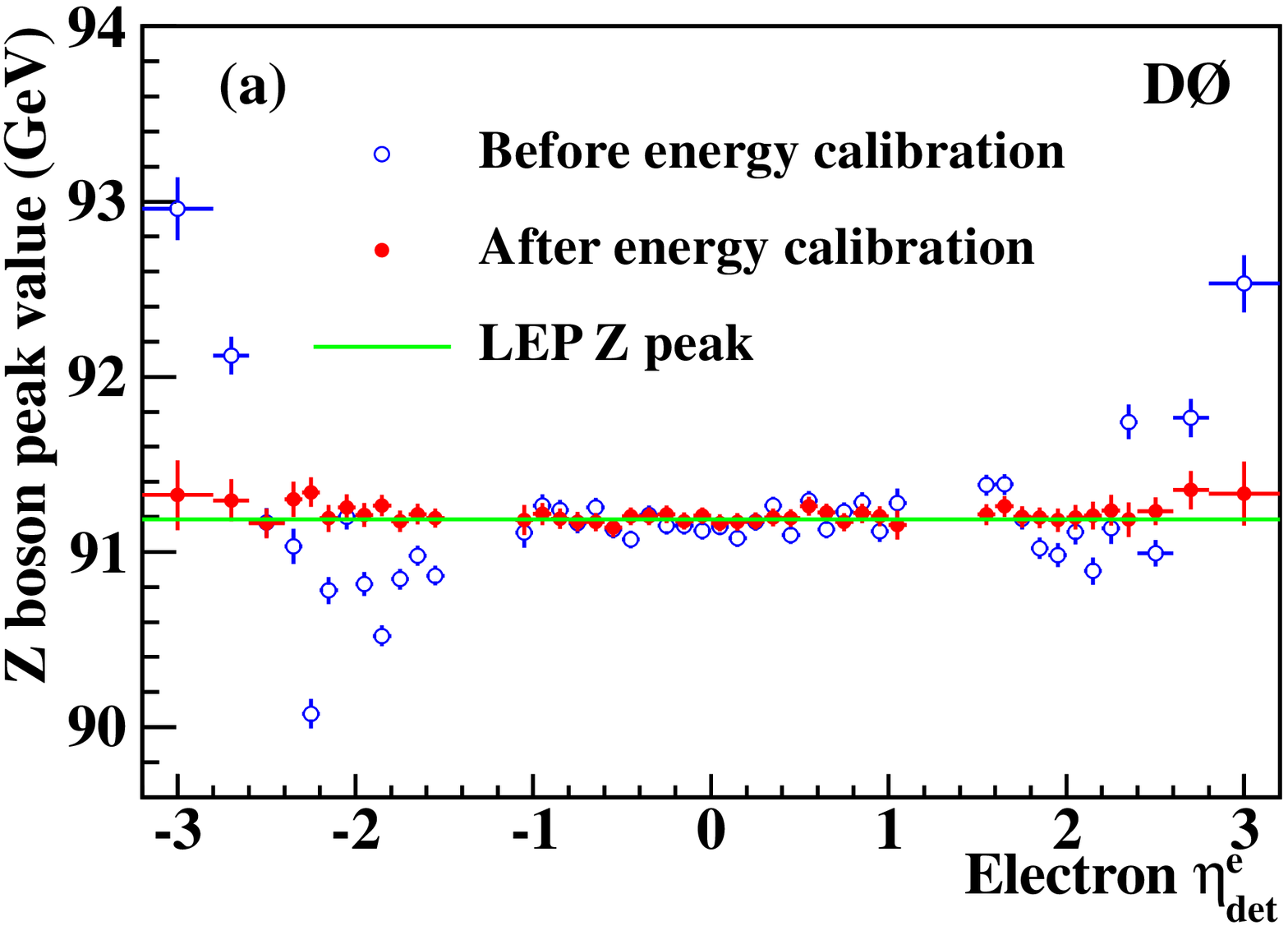}}
\end{minipage}
\caption{Charge mis-ID for MC and data (left) and the fitted mass value of $Z \rightarrow ee$ data events where one electron is central and the other is forward (right).}
\label{fig:calib}
\end{figure}

The charge mis-ID is determined using a tag and probe method with a $Z \rightarrow ee$ sample, as a function of electron $\eta$ and $p_{T}$.  The efficiency is shown in Figure~\ref{fig:calib}, left.  The Monte Carlo (MC) and data particularly disagree in the forward region where the mis-ID probability is high.  Because the MC does not accurately reflect the data, the MC charge is randomly flipped until the mis-ID values match, to adjust for mis-modeling.

The electron energy scale and offset are determined using a background subtracted $Z \rightarrow ee$ sample.  These events are fit to determine the \Z mass peak, which is compared to the LEP value of 91.1876 GeV~\cite{lep}.  Correction parameters are then determined iteratively.  The calibration is performed as a function of electron $\eta$, luminosity, and calorimeter scalar $E_{T}$.  Figure~\ref{fig:calib}, right, shows the agreement with the LEP value before and after the calibration.  The improvement in the central region is small, where the data already largely agreed with the LEP value.  However, in the forward region the energy calibration is particularly impactful.

\section{Lepton Method}
We can approximate the asymmetry as simply the difference in the number of charged electrons over the sum, with respect to electron pseudorapidity.  Effects of electron selection efficiencies, luminosity, and event acceptance on the number of electrons are all accounted for and the analysis is unfolded, removing detector effects to allow comparison with the predictions.  Details are available in the analysis PRD.  Because of CP symmetry, positive and negative $\eta$ regions have equivalent asymmetry and the data from these two regions are combined.

\begin{figure}[htp]
  \centering
  \begin{tabular}{cc}
    \includegraphics[width=70mm]{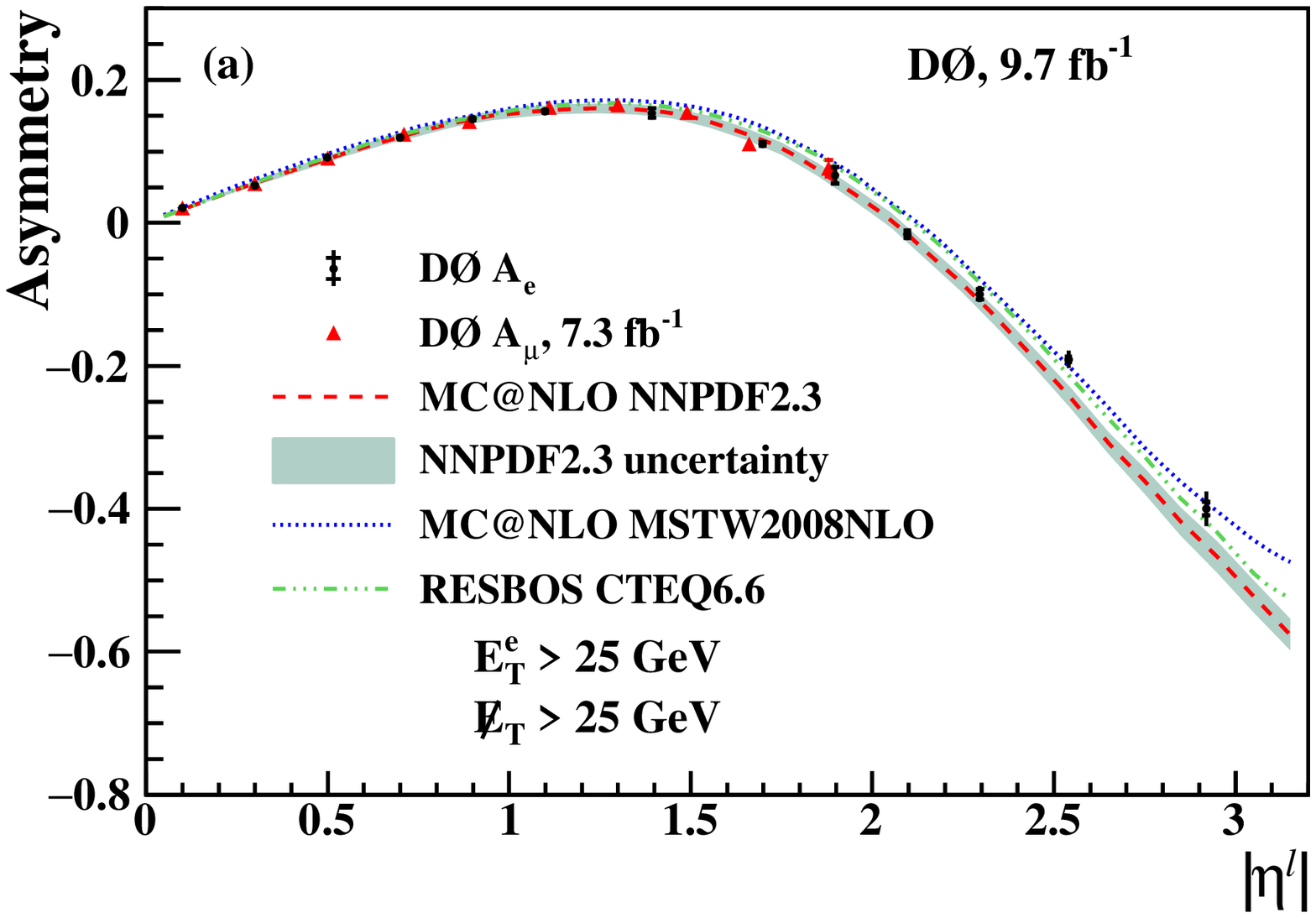}&
    \includegraphics[width=70mm]{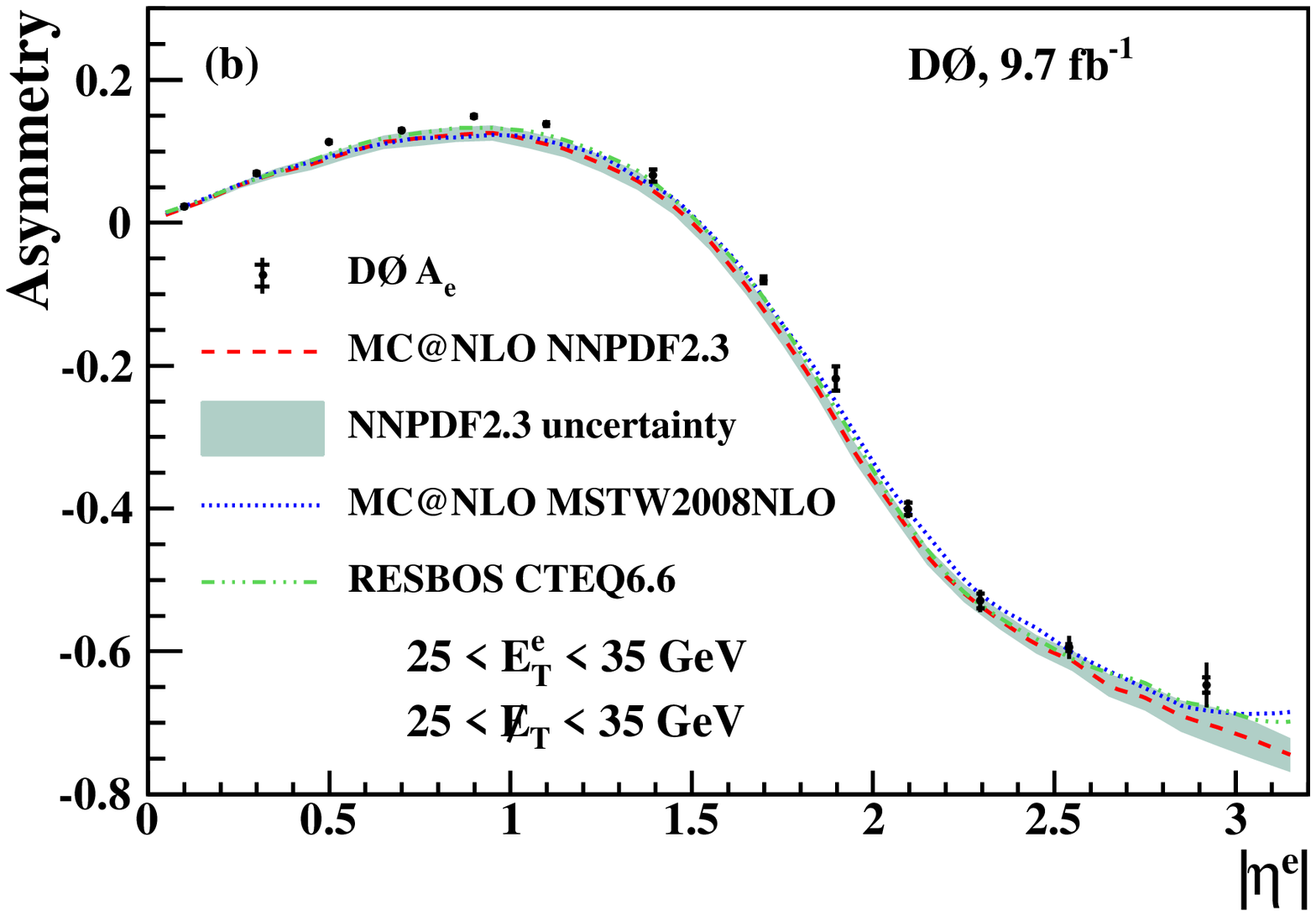}\\
    \includegraphics[width=70mm]{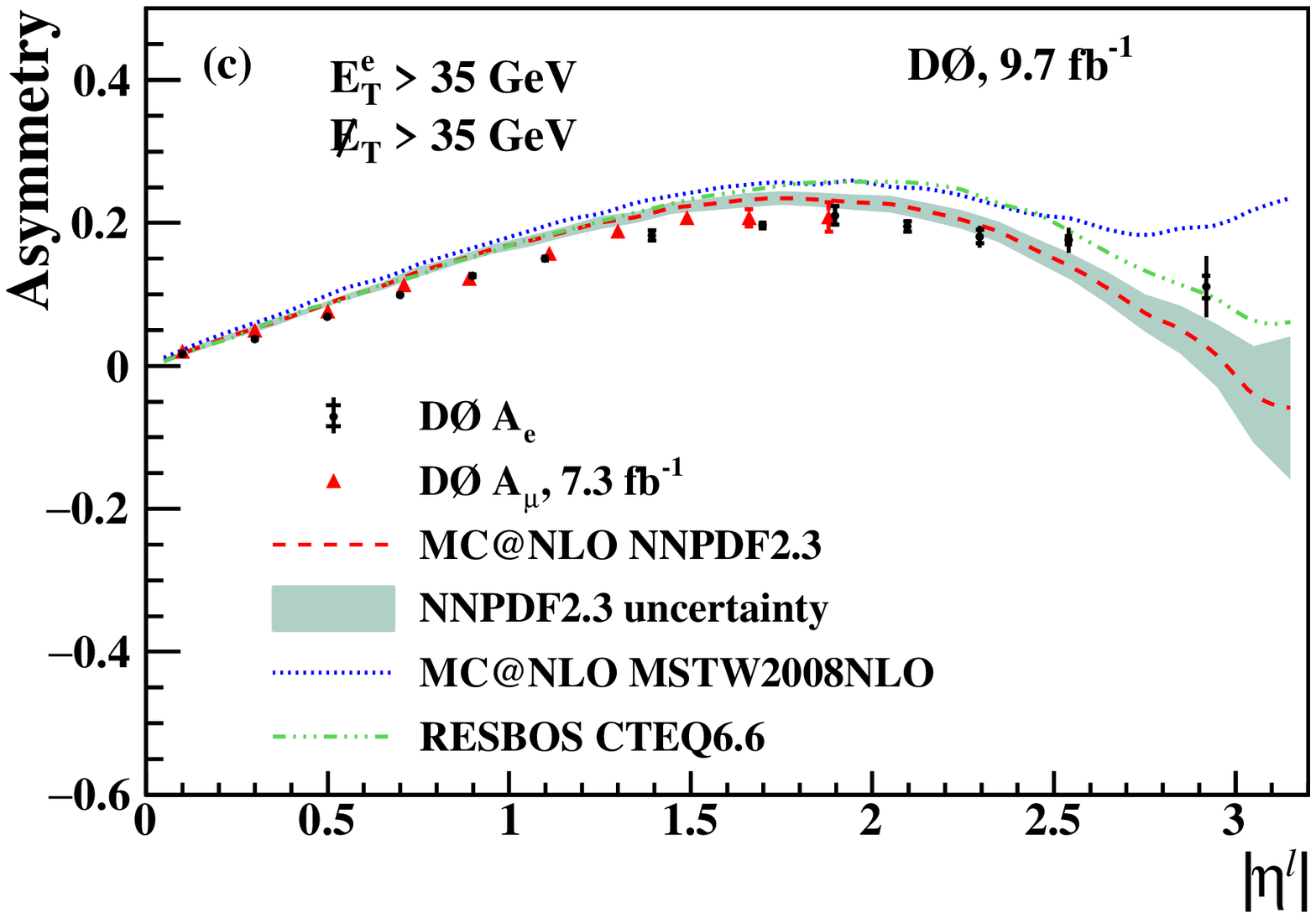}&
    \includegraphics[width=70mm]{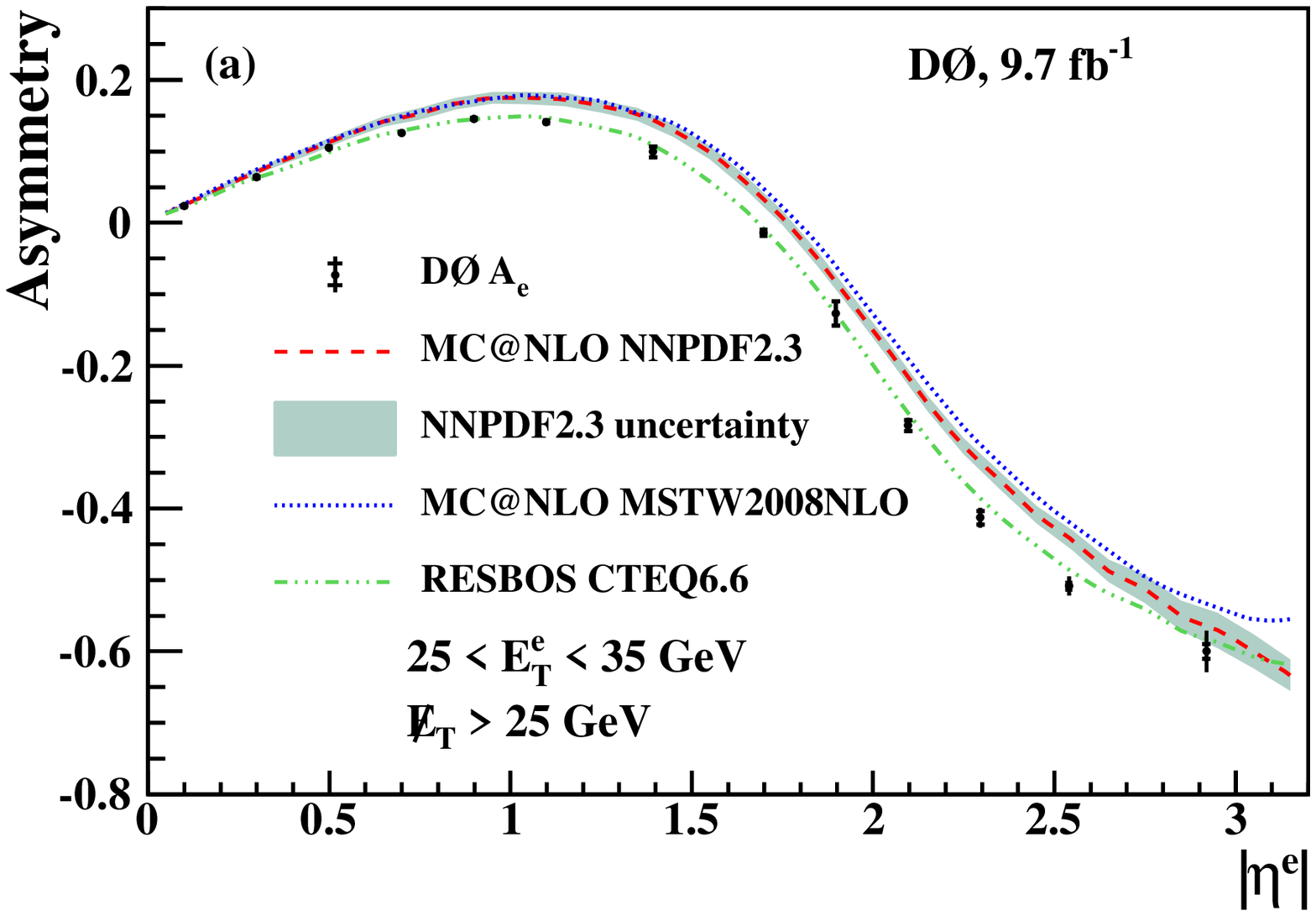}\\
    \includegraphics[width=70mm]{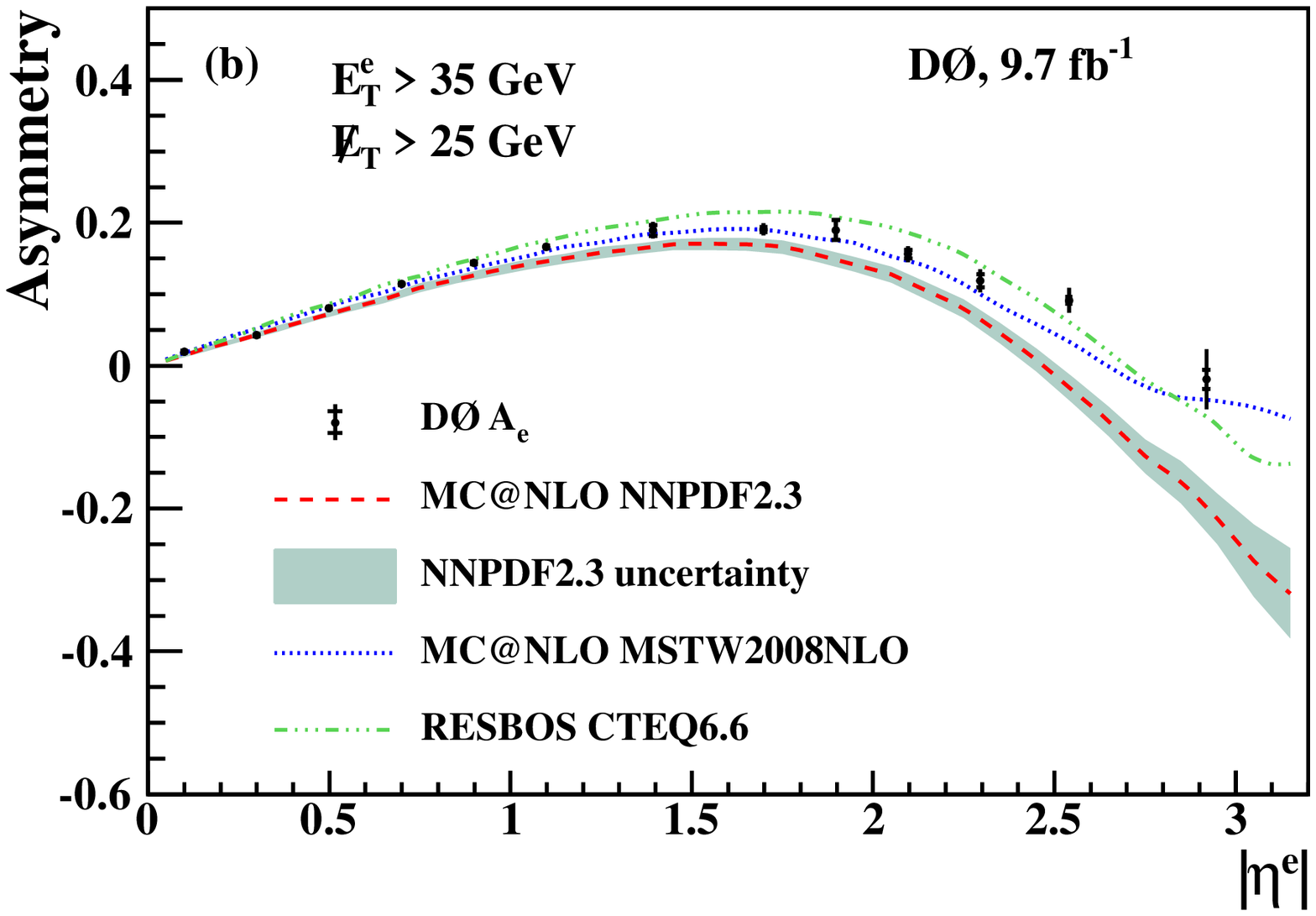}&
    \\
  \end{tabular}
  \label{fig:lepasym1}\caption{Measured \W boson charge asymmetry using the lepton rapidity method showing the result from this work using electrons, the previous 7.3\fb D0 result using muons, and various predictions listed in the legend. Values are given after CP folding.  Vertical lines show the total uncertainty, and horizontal lines indicate the statistical component.}
\end{figure}

The results are reported as a function of lepton $\eta$ and in symmetric and asymmetric bins of \met and lepton $p_{T}$.  These bins, and corresponding distributions, are shown in Figure~\ref{fig:lepasym1}.  The figure shows comparisons with a previous muon channel D0 analysis and various predictions, namely MC@NLO~\cite{mcatnlo} with NNPDF2.3~\cite{nnpdf}, NLO {\sc resbos}~\cite{resbos} plus {\sc photos}~\cite{photos} with CTEQ6.6~\cite{cteq}, and MC@NLO with MSTW2008NLO~\cite{mstw}.  There is also supplementary material available with the PRD for this study showing an additional dataset comparison, which is not discussed here.  Correlation coefficient matrices for this result are also available.  Overall, the predictions agree well with the data in the first bin (\met > 25 GeV, electron \pt > 25 GeV).  In the other bins, the agreement with CTEQ6.6 is good for the asymmetric bins but all predictions diverge from the data in the two symmetric bins.

\section{Reconstructed \W Method}
To reconstruct the W and report the rapidity, we need to fully reconstruct the neutrino momentum.  Although the neutrino \pz is not a measured quantity at hadron colliders, it can nevertheless be estimated using the \W boson mass (80.385 GeV~\cite{wmass}) and other event information.  The most likely \pz is then used to reconstruct the \W boson (particularly its rapidity).  The W boson mass is related to the sum of the squares of its final state electron and neutrino energy and final state momenta, which allows the determination of the neutrino \pz to within a two-fold ambiguity.  In the case of a complex result, a real solution is always obtained by assuming the \met was mis-reconstructed and adjusting the \met until the result is real.  The ambiguity between two equation results is resolved by determining weights for the event, for each solution, using cos$\theta^*$, \W rapidity and \W \pt information.  The \W information used in the weight is obtained from generators, and the weights are updated in an iterative way, to remove any analysis bias, until the weights converge.  Details regarding this method may be found elsewhere~\cite{wasymmethod,D0:wasym,CDF:wasym}.

\begin{figure}
\begin{minipage}{0.50\linewidth}
\centerline{\includegraphics[width=0.9\linewidth]{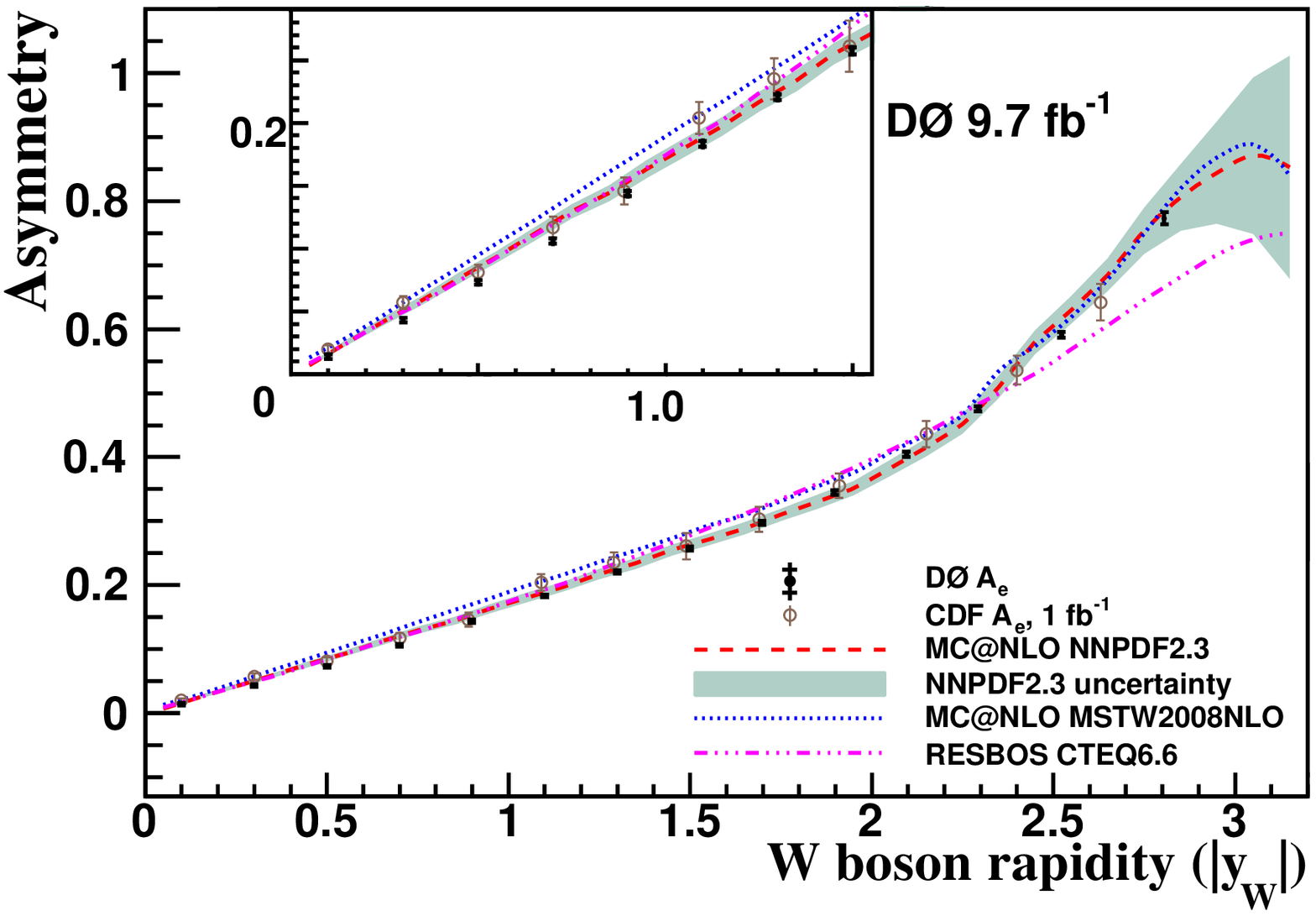}}
\end{minipage}
\hfill
\begin{minipage}{0.50\linewidth}
\centerline{\includegraphics[width=0.9\linewidth]{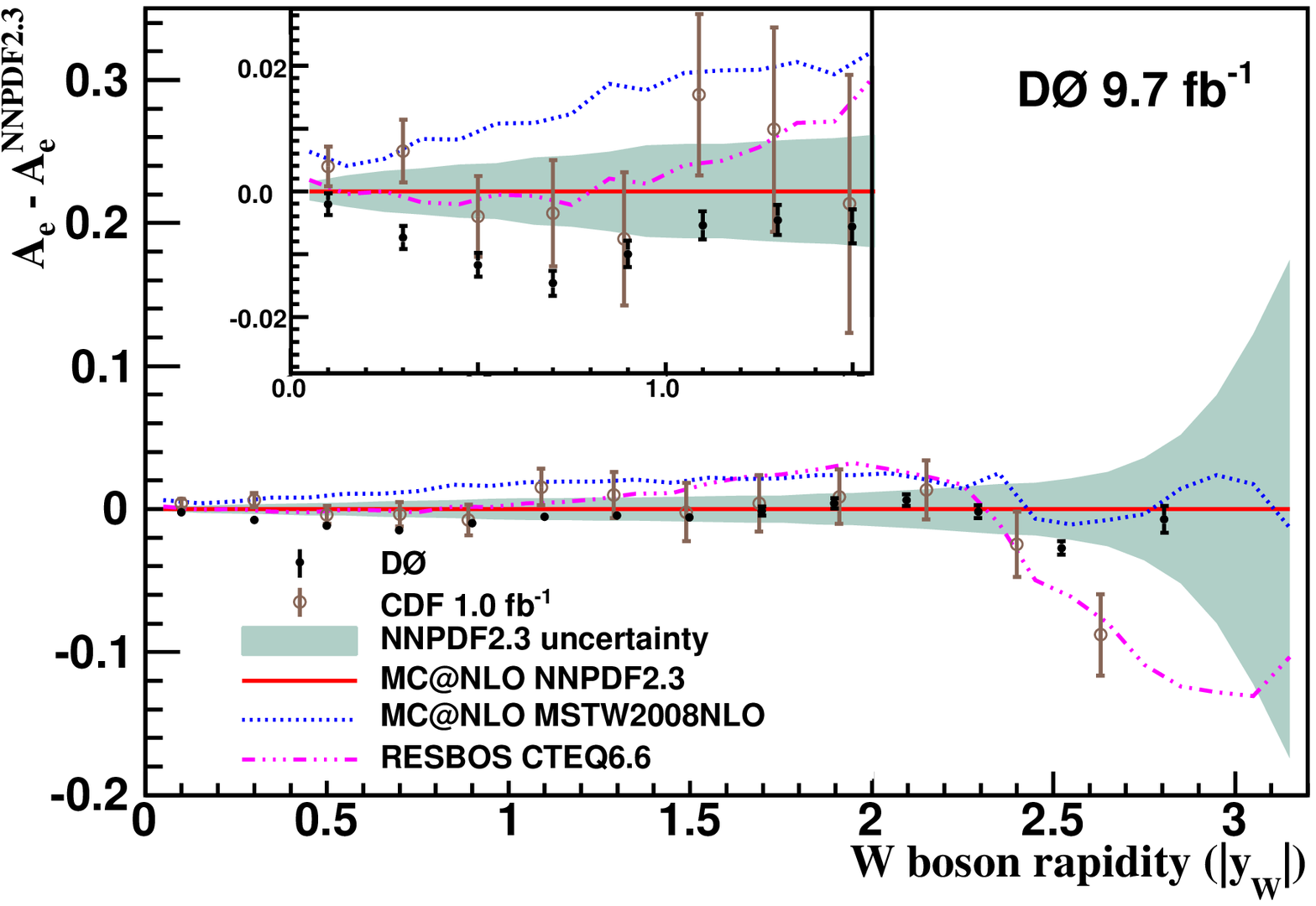}}
\end{minipage}
\caption{Measured \W boson charge asymmetry using the \W rapidity method (left) and with respect to NNPDF 2.3 (right) showing the result from this work, the previous 1.0\fb CDF result, and various predictions listed in the legend.  The inset shows the \W rapidity region from 0 to 1.5.  Values are given after CP folding.  Vertical lines show the total uncertainty, and horizontal lines indicate the statistical component.}
\label{fig:Wasym}
\end{figure}

As with the previous method, the analysis is unfolded, and positive and negative $\eta$ regions are combined.  The result is reported in one inclusive bin of \met and lepton $p_{T}$, given in Figure~\ref{fig:Wasym}.  The data are compared with the same predictions as the previous method, and with a previous CDF result.  A correlation coefficient matrix for this result is also available.  Overall the agreement with the various predictions is reasonable, as is the agreement with the CDF data result.  The predictions are a bit higher than the data in the central rapidity region and overall the data uncertainties are lower than the given prediction uncertainty, indicating the usefulness of this data for future PDF sets.

\section{Summary}
The measurement of the \W boson asymmetry using data from the Fermilab Tevatron Collider is a particularly important input to various future PDF set fits.  We have reported recent measurements using the full D0 data set, using two different methods, which make use of the electron $\eta$ or reconstructed \W rapidity distributions.  Both the lepton and \W rapidity versions of this measurement are the most precise to date.

\end{document}